\documentclass[twocolumn,prb,aps,showpacs]{revtex4-1}
\usepackage{graphicx}
\usepackage{dcolumn}
\usepackage{amsmath}
\usepackage{units}
\usepackage[utf8]{inputenc} 
\usepackage{kotex}

\begin{document}
\title{A weak topological insulator state in quasi-one-dimensional superconductor TaSe$_3$}
\author{Jounghoon Hyun$^{1\dagger}$, Min Yong Jeong$^{1\dagger}$, Sunghun Kim$^1$, Myung-Chul Jung$^1$,\\ Yeonghoon Lee$^1$, Chan-young Lim$^1$, Jaehun Cha$^1$, Gyubin Lee$^1$, Yeojin An$^1$,\\ Makoto Hashimoto$^2$, Donghui Lu$^2$, Jonathan D. Denlinger$^3$, Myung Joon Han$^{1*}$ and Yeongkwan Kim$^{1,4*}$}
\address{$^1$Department of Physics, Korea Advanced Institute of Science and Technology, Daejeon 34141, Korea.}
\address{$^2$Stanford Synchrotron Radiation Lightsource, Stanford Linear Accelerator Center, Menlo Park, CA 94025, USA.}
\address{$^3$Advanced Light Source, Lawrence Berkeley National Laboratory, Berkeley, CA 94720, USA.}
\address{$^4$Graduate School of Nanoscience and Technology, Korea Advanced Institute of Science and Technology, Daejeon 34141, South Korea.}

\maketitle

\vspace{10 pt}

{\bf 
A well-established way to find novel Majorana particles in a solid-state system is to have superconductivity arising from the topological electronic structure. To this end, the heterostructure systems that consist of normal superconductor and topological material have been actively explored in the past decade. However, a search for the single material system that simultaneously exhibits intrinsic superconductivity and topological phase has been largely limited, although such a system is far more favorable especially for the quantum device applications. Here, we report the electronic structure study of a quasi-one-dimensional (q1D) superconductor TaSe$_3$. Our results of angle-resolved photoemission spectroscopy (ARPES) and first-principles calculation clearly show that TaSe$_3$ is a topological superconductor. The characteristic bulk inversion gap, in-gap state and its shape of non-Dirac dispersion concurrently point to the topologically nontrivial nature of this material. The further investigations of the Z$_2$ indices and the topologically distinctive surface band crossings disclose that it belongs to the weak topological insulator (WTI) class. Hereby, TaSe$_3$ becomes the first verified example of an intrinsic 1D topological superconductor. It hopefully provides a promising platform for future applications utilizing Majorana bound states localized at the end of 1D intrinsic topological superconductors.
}

Topological superconductors are recently attracting great research attention due to their fundamental importance and potential application to the quantum computation based on the Majorana bound states\cite{Kitaev_2001, Leijnse_2012, RevModPhys.83.1057, doi:10.7566/JPSJ.85.072001, Sato_2017}. A popular approach is to make use of superconducting proximity effect\cite{Alicea_2012} with which Cooper pairs are injected into the topological surface state as theoretically suggested by Fu and Kane\cite{PhysRevLett.100.096407}. Following this idea, several heterostructures have been fabricated and indeed shown to host the Majorana quasiparticles\cite{wang_coexistence_2012, mourik_signatures_2012, Jack:2019aa}. From the point of view of the application, however, the heterostructure systems can likely bring technical difficulties in the device fabrication. It is therefore highly desirable to have a material that intrinsically hosts topological superconductivity, although not many systems have been experimentally confirmed to be the case. One promising candidate is Fe(Te,Se) where superconductivity is encoded into the topological surface state on the two-dimensional (2D) surface\cite{Zhang182}. In this material, the topological superconductivity can certainly be intrinsic. However, the location of the Majorana state can be at the center of the vortex or anywhere on the edge of 2D surface; namely, it is not pre-determined locally.

In this regard, an important research direction, that has not been well explored so far, is to search for intrinsic topological superconductivity in one-dimensional (1D) systems. Different from the 2D or three-dimensional (3D) case, Majorana states exist at both ends of the 1D topological superconductor\cite{Nadj-Perge602, Jack:2019aa}, thus making its manipulation much easier. It would be useful for circuit device applications by securing the Majorana states at certain desired positions. Here we note a recent theoretical study suggesting that TaSe$_3$ is a strong topological insulator (STI)\cite{nie_topological_2018}. Also, the previous experiment seem to indicate the unusual superconductivity in this material by showing the absence of a diamagnetic response below the critical temperature $T_\mathrm{C}$\cite{NAGATA1991761}. To be the first confirmed example of 1D intrinsic topological superconductivity, however, further investigations of its topological properties are strongly requested from both experimental and theoretical sides. 

In this Communication, we report our detailed study of the electronic structure of TaSe$_3$ by using angle-resolved photoemission spectroscopy (ARPES) and first-principles density functional theory (DFT) calculation. The band inversion by spin-orbit coupling (SOC) and the additional in-gap surface state with non-Dirac dispersion clearly indicate the nontrivial topology of its band structure. Further analyses of Z$_2$ indices and the number of Dirac points depending on the surface type clearly show that TaSe$_3$ is a weak topological insulator (WTI). Our results establish TaSe$_3$ as the first experimentally verified example of an intrinsic 1D topological superconductor.\\

\begin{figure*}
	\centering\includegraphics[width=16.5 cm]{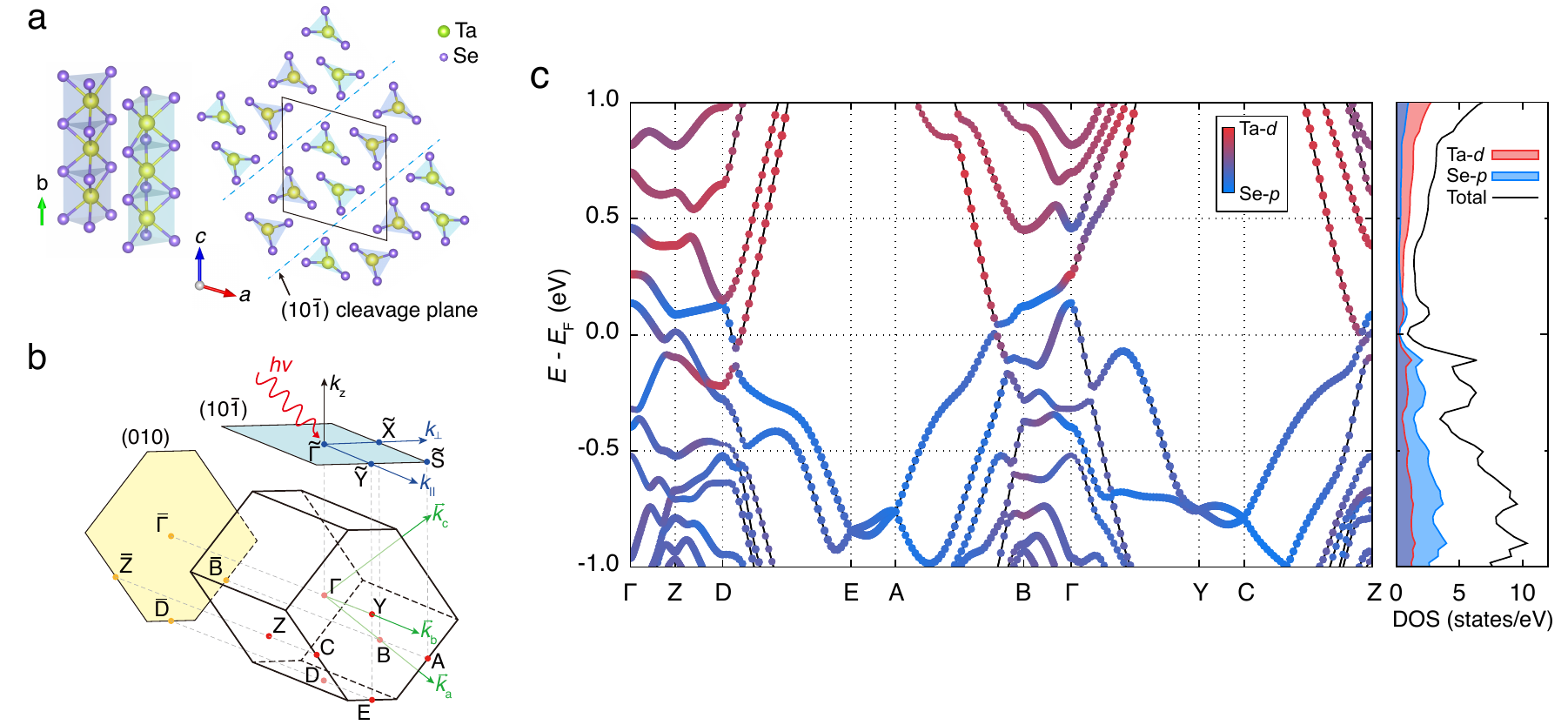}
	\caption{{\bfseries Crystal structure and calculated bulk band structure of TaSe$_3$.} {\bfseries a}, Crystal structure of TaSe$_3$. Ta and Se atoms are depicted by yellow and violet spheres, respectively. The black solid parallelogram and blue dashed line shows the unit cell and the practical cleavage plane (10\={1}). {\bfseries b}, 3D and surface-projected BZs. The planes colored by blue and yellow represent the surface BZs of (10\={1}) and (010) planes, respectively. $k_\parallel$ and $k_\perp$ indicate projected reciprocal lattice vectors on the (10\={1}) surface BZ. {\bfseries c}, Calculated bulk band structure (left) and projected density of states (PDOS, right). The projected weights of Ta-5$d$ and Se-4$p$ orbitals are presented in red and blue colors, respectively. The band inversions are clearly observed around the high-symmetry points of B, Z, and D. For the calculated band structure without SOC, see Supplementary Fig. 2.}
    	\label{fig1}
\end{figure*}

\noindent\textbf{Results}

\noindent\textbf{Crystal and electronic structure.}
TaSe$_3$ consists of elongated prismatic chains with monoclinic P2$_1$/m space group symmetry\cite{Bjerkelund_1965,Bjerkelund_1966} (Fig. 1a), which gives rise to the quasi-one-dimensional (q1D) electronic properties. The chains are stacked via van-der-Waals interaction and the weak potential modulation along with the chain-perpendicular direction results in the finite band dispersion within 3D Brillouin zone (BZ) (Fig. 1b). Due to the different bonding strengths between interchain Se ions, (10\={1}) plane becomes a natural cleavage plane on which ARPES measurements are conducted. Figure 1b shows the eight time-reversal invariant momentum (TRIM) points and their projections onto the surface BZs. Hereafter, we will use the notations of the high-symmetry points in the surface-projected BZ and in 3D BZ for describing ARPES data and bulk band calculations, respectively.

Figure 1c presents the calculated bulk band structure of TaSe$_3$. The low-energy states are governed by Ta-5$d$ and Se-4$p$ orbital characters and their relative weights are depicted by red and blue color, respectively. The sizable hybridization between these two orbitals are clearly noted in the projected density of states (PDOS; right panel of Fig. 1c). The orbital-weighted band structure shows the band inversions around B, Z, and D points, which indicates its topological nature. It becomes clearer in comparison to the calculated band structure without SOC (see Supplementary Fig. 2). The SOC removes the band crossings and induces the gap openings accompanied by band inversion; see the band structure changes along e.g., D--E, B--A and Z--C line.\\

\begin{figure*}[ht]
	\includegraphics[width=16.5 cm]{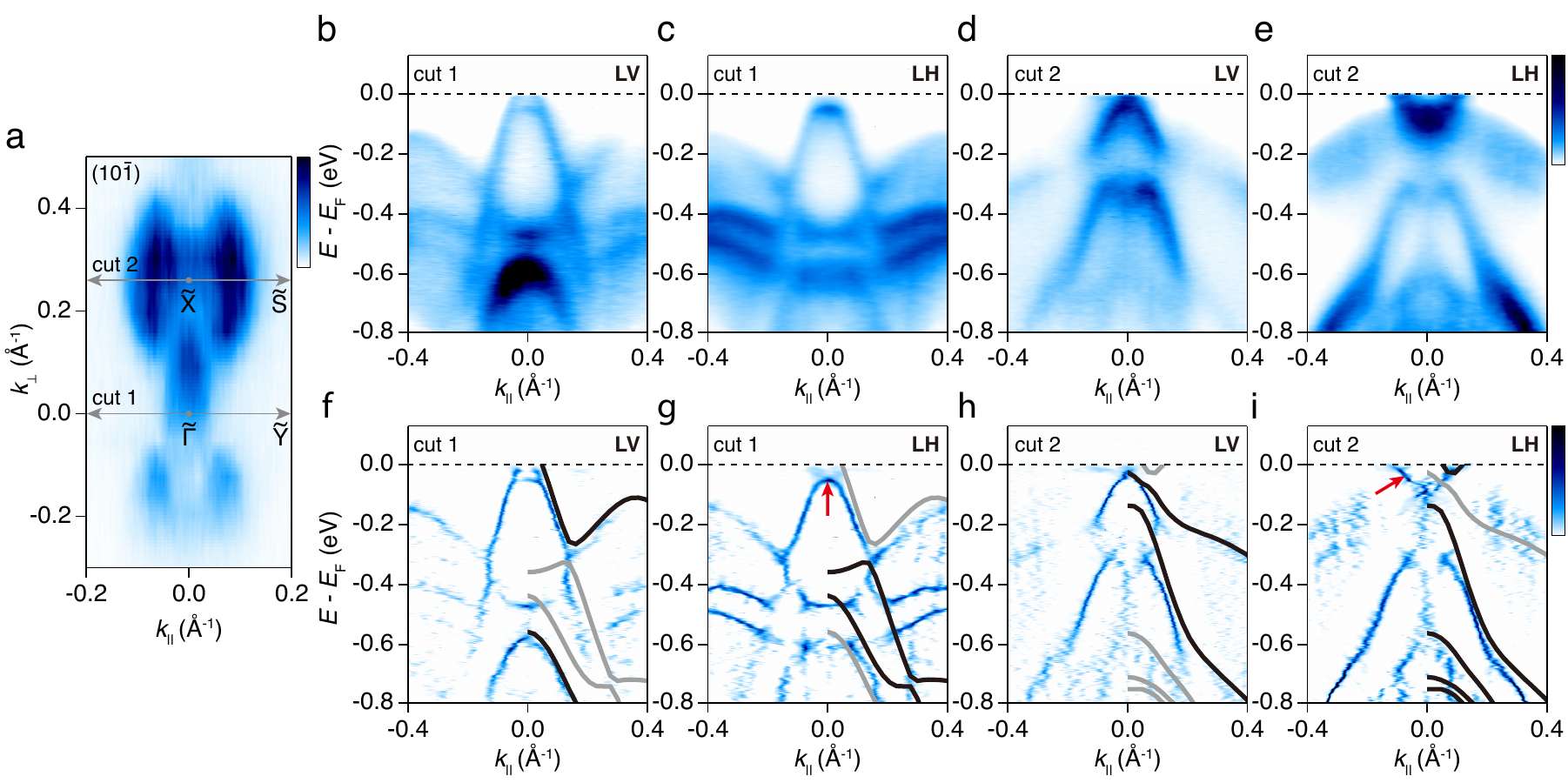}
	\caption{{\bfseries ARPES results of TaSe$_3$ in comparison to the calculated bulk band structure.} {\bfseries a}, Fermi surface measured at (10\={1}) plane. $k_\parallel$ ($k_\perp$) refers to the direction parallel (perpendicular) to the chain. Gray solid lines with arrows indicate two high symmetry lines,  `cut 1' (\~{Y}--{$\tilde{\Gamma}$}--\~{Y}) and `cut 2' (\~{S}-\~{X}-\~{S}), along which further investigations were carried out. {\bfseries b-c}, ARPES spectra along cut 1, obtained by utilizing LV and LH polarized light. {\bfseries d-e}, ARPES spectra along cut 2, obtained by utilizing LV and LH polarized light. {\bfseries f-i}, Curvature plots of {\bfseries (b-e)}. The calculated bulk band structures are superimposed on the right half of the plots for comparison. Black solid lines (gray solid lines) indicate the visible (invisible) bands in each polarization. The red arrows in \textbf{(g,i)} indicate the bands which are observed by ARPES but not found in the bulk band calculation.}
\label{fig2}
\end{figure*}

\noindent\textbf{ARPES electronic structure and an in-gap state.} In order to examine the electronic structure and its topological nature, we carried out ARPES measurement on high-quality single crystals of TaSe$_3$. Figure 2 presents the ARPES spectra measured on the cleaved (10\={1}) surface. A constant energy contour plot at Fermi level ($E\mathrm{_F}$) (Fig. 2a) shows an elongated Fermi surface along $k_\perp$, reflecting the overall 1D character of this material ($x$- and $y$-axis corresponds to $k_\parallel$ and $k_\perp$ direction in Fig. 1b, respectively). The band dispersions were measured along two projected high-symmetry lines, namely, {$\tilde{\Gamma}$}--\~{Y} (cut 1) and \~{X}--\~{S} (cut 2). These two lines are particularly interesting because the characteristic band inversions are observed (see Fig. 1b, c) and the Dirac cone, a hallmark of a topological insulator, has been predicted by the previous calculation\cite{nie_topological_2018}.

In taking the ARPES spectra, we utilized both linear vertically (LV) and horizontally (LH) polarized incident light to avoid missing of the spectral weight caused by matrix element effect. Note that each spectrum taken with different polarizations highlights different bands and is therefore complementary to each other. For instance, two hole bands with different band-top energies near $E\mathrm{_F}$ are well identified in the cut 1 taken with LV (Fig. 2b) while only the lower one is seen by LH (Fig. 2c). Along cut 2, LV clearly reveals a hole band near $E\mathrm{_F}$ (Fig. 2d) whereas LH proves the relatively large electron-like band together with additional two tiny electron bands which barely cross the $E\mathrm{_F}$ (Fig. 2e). To analyze the complicated band dispersion of TaSe$_3$ more clearly, the curvature method \cite{doi:10.1063/1.3585113} is applied (Figs. 2f-i) which enables us to make more direct comparisons with calculation results. 

As shown in Fig. 2b-i, the overall band structure of ARPES spectra is in good agreement with our theoretical result. The calculated bulk bands along $\Gamma$--Y (Z--C) are presented as solid lines and overlaid on the curvature plots of cut 1 (cut 2) in Fig. 2f-i where the black (gray) lines represent the calculated bands that are observed (not observed) in the corresponding polarization light of ARPES data. From the point of view of topology, most intriguing is the tiny electronic bands whose bottom is located just below $E\mathrm{_F}$; see the small parabolic band (solid black line) in Fig. 2i (it is not visible with LV polarization and depicted by the gray line in Fig. 2h). This band is symmetric with respect to $k_{\parallel}=0$ (\~{X}) as seen in our curvature plot; see the left half of Fig. 2h-i (not overlaid with the calculation result of a solid line). Note that the bottom of this band is not located at \~{X}. This characteristic feature caused by band inversion reflects its topological nature as discussed recently by Tang {\it et al.}\cite{Tang:2017aa}. Thus, our result provides the first experimental signature of nontrivial topology of this q1D superconductor.

Importantly, some noticeable band features in ARPES spectra are not observed in the calculation. See, in particular, two bands near $E\mathrm{_F}$ indicated by red arrows in Figs. 2g and 2i. The first one is the two hole bands in cut 1 (Fig. 2g). While the upper-hole band is well identified in our calculation (see the uppermost black solid line in Fig. 2h), the lower is only observed in ARPES. The second is a small piece of the electron-like band in cut 2 locating in between the lower-lying large hole bands and the upper tiny electron bands; see Fig. 2i. Once again, the spectrum of this band is large enough in ARPES but not seen in DFT result. Understanding the discrepancy between theory and experiment, particularly for the case of this second band, is of key importance because it locates in the band inversion gap and thus is likely to be involved with the topological nature of this material. In the below, we will show that this `in-gap state' has in fact the surface origin and its dispersion feature can be regarded as an indication of WTI state rather than STI\cite{nie_topological_2018}.\\

\begin{figure*}[ht]
	\centering\includegraphics[width=16.5 cm]{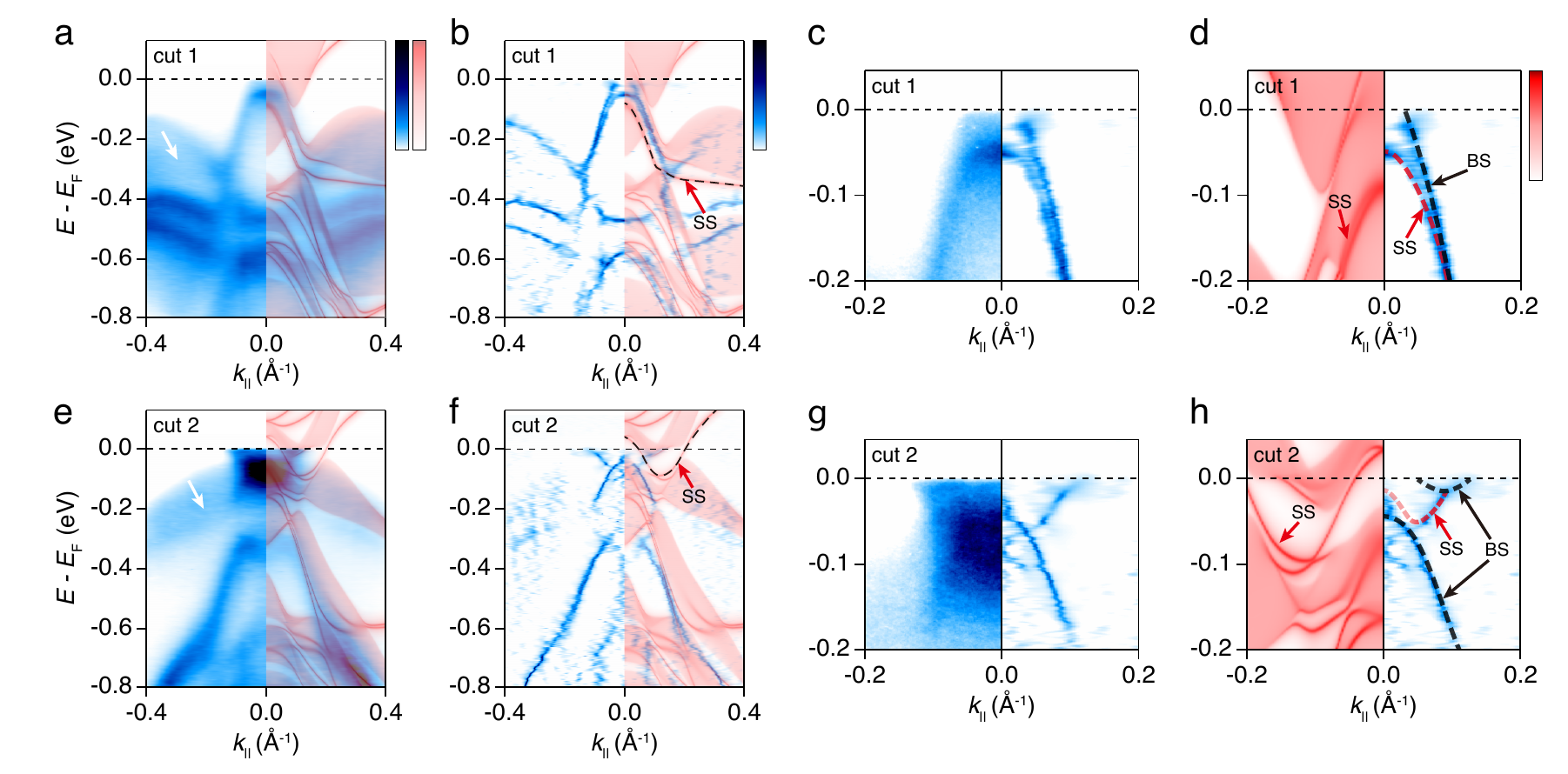}
	\caption{{\bfseries The (10\=1) surface states revealed by ARPES and the slab calculation.} {\bfseries a-d}, ARPES spectra along cut 1 (\~{Y}--{$\tilde{\Gamma}$}--\~{Y}) and its curvature plot in comparison with the corresponding slab band calculation. {\bfseries a}, ARPES spectra along cut 1 obtained by adding up the LV and LH data (Figs. 2b, c). The slab calculation results are overlaid in the right half of the figures. The white arrow indicates the broad continuum state, which is consistently found in both experiment and calculation. {\bfseries b}, Curvature plot of {\bfseries (a)} overlaid by the same slab calculation. The black dashed line shows the surface-originate state (`SS').
 {\bfseries c}, Zoomed-in ARPES image of {\bfseries (a)} (left) and {\bfseries (b)} (right). {\bfseries d}, Zoomed-in image of {\bfseries (b)}. The black dashed line (`BS') represents the band from the bulk band calculation (guide to the eye) and the red dashed line (`SS') indicates the band not observed in the bulk calculations. The surface band spectra are well separated from the nearby bands. {\bfseries e-h}, The same with {\bfseries (a-d)} along cut 2 (\~{S}-\~{X}-\~{S}). {\bfseries h}, Overlaid red dashed line represents the in-gap state observed in ARPES spectra which is missing in the bulk band calculation. It is also clearly observed in the left half of the panel as indicated by the red arrow.}
	\label{fig3}
\end{figure*}

\noindent \textbf{Surface electronic structure and band topology: DFT calculation.} 
To further elucidate the nature of band dispersion including the in-gap state, we performed the slab calculation by using iterative Green's function method\cite{sancho_highly_1985,wu_wanniertools:_2018} based on maximally localized Wannier functions (MLWFs)\cite{marzari_maximally_1997,souza_maximally_2001,mostofi_wannier90:_2008} generated from DFT bands (see Methods for more details). In Fig. 3, the results are presented and compared with experimental spectra where two ARPES spectra taken from both polarizations are added up. Besides the overall good agreement in between calculation and experiment, we first take a special note on the broad continuum states (indicated by white arrows in Figs. 3a, e). This continuum is well identified also in the calculation being attributed to the surface projection of bulk states at a range of $k\mathrm{_z}$ values (see Fig. 1b). In ARPES measurement, on the other hand, the broadening of photoelectron momentum along $k\mathrm{_z}$ direction is due to the finite probing depth of incident light in real space. Therefore, this consistency between theory and experiment confirms that our slab calculation sufficiently well describes the states near the surface.

Interestingly, Fig. 3 clearly shows that the two aforementioned bands (which are present in ARPES but not observed in the bulk calculation) are  originated from surface states. For clearer comparisons, see Fig. 3b and 3f for the case of hole-like (in cut 1; {$\tilde{\Gamma}$}--\~{Y} line) and electron-like band (in cut 2; \~{X}--\~{S} line), respectively, as denoted by the black-dashed lines and the red arrows. From the fact that these bands are only reproduced by surface band calculations (and not found in 3D bulk calculations), we conclude that they are the surface states. Another fact that these bands are well separated from the nearby bulk spectra also supports this conclusion.

From the point of view of band topology, important is the unambiguous identification of the second band (i.e., the electron-like band indicated by `SS' in Figs. 3f, h) as being originated from the surface. As is well known, a hallmark of topological insulator is the characteristic surface state with Dirac crossing, which is located inside the inversion gap\cite{fu_topological_2007}. The absence of Dirac cone at $(10\bar{1})$ surface leads us to conclude that TaSe$_3$ is not STI contrary to the previous theoretical suggestion\cite{nie_topological_2018}, and to require  further investigation regarding its topological nature.\\

\begin{figure*}[ht]
	\centering\includegraphics[width=16.5 cm]{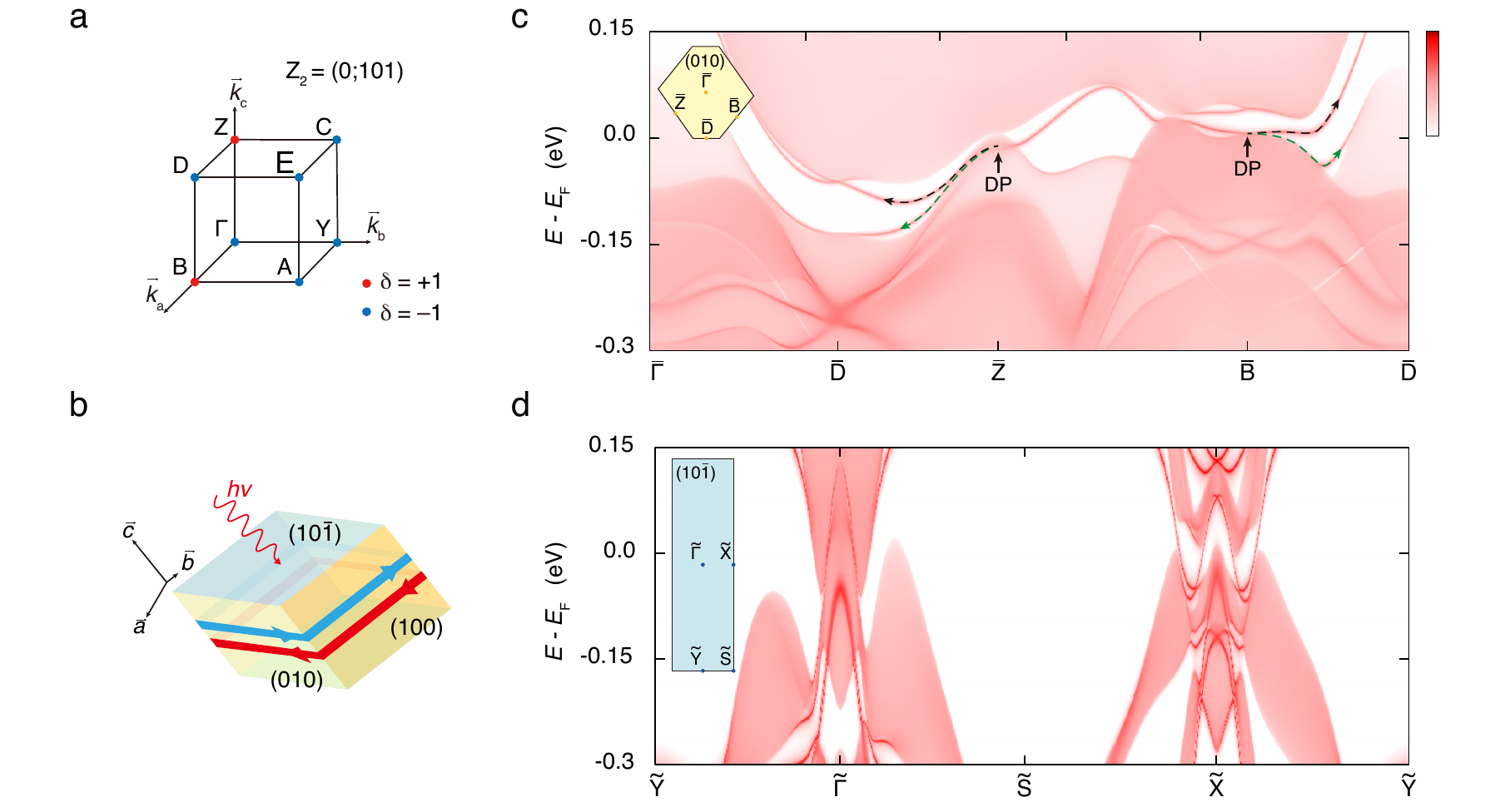}
	\caption{{\bfseries The calculated Z$_2$ topological invariants and the surface Dirac points.} {\bfseries a}, The products of band parities for all occupied bands ($\delta$) at eight bulk TRIM points. The red and blue dots represent $\delta=+1$ and $-1$, respectively. {\bfseries b}, Schematic picture for the TaSe$_3$ surfaces and their topological nature. According to the $Z_2$ classification\cite{fu_topological_2007}, the (10\={1}) surface (blue colored) should be topologically dark while two other side surfaces, (010) and (100), are expected to have two Dirac points on each plane and to host topologically protected helical surface states (depicted by blue and red arrow). {\bfseries c},  The calculated (010) surface band structure. The inset shows the surface-projected BZ and the relevant high-symmetry points. Two Dirac points are found as expected at $\bar{\rm{B}}$ and $\bar{\rm{Z}}$ points. The topological surface states are indicated by the black (connecting to the conduction band) and the green dashed lines (connecting to the valence band) with arrows. The color bar represents the intensity strength. {\bfseries d} The same as in {\bfseries (c)}, for the $(10\bar{1})$ surface. As expected, no Dirac cone is found.}
	\label{fig4}
\end{figure*}

\noindent \textbf{Discussion}

\noindent \textbf{$\mathbf{Z_2}$ topological indices and weak topological insulator.} According to Fu {\it et al.}\cite{fu_topological_2007}, STI has surface Dirac cones on its surface. On the other hand, WTI has Dirac surface state only on a certain type of surfaces. The other type of surface is called topologically `dark'. Here it is important to note that the cleaving surface of TaSe$_3$ is in fact a dark surface. Therefore, our observation of no Dirac point in ARPES spectra does not exclude the intriguing possibility for TaSe$_3$ to be the first confirmed example of q1D topological superconductor. In order to examine the possibility of WTI, we calculated Z$_2$ topological indices, and found that TaSe$_3$ is indeed a WTI;  ($\nu_{0}$; $\nu_{1}$, $\nu_{2}$, $\nu_{3}$)=(0;101). Figure 4a shows the products of the occupied band parities at the eight TRIM points from which ($\nu_{0}$; $\nu_{1}$, $\nu_{2}$, $\nu_{3}$) are defined\cite{fu_topological_2007}. As is well established, $\nu_{0}$ = 0 and $\nu_{1,3}\neq 0$ are a clear signature of WTI\cite{fu_topological_2007}.

In order to further confirm the WTI nature of TaSe$_3$, we explored the other surfaces than (10$\bar{1}$), namely the `non-dark' surfaces on which the surface Dirac cones are expected to appear. The dark surface of WTI is defined by the Miller indices of the given surfaces. That is, the surfaces defined by ($\nu_{1}$, $\nu_{2}$, $\nu_{3}$) and ($\nu_{1}$, $\nu_{2}$, $\nu_{3}$) + 2$\vec{G}$ ($\vec{G}$: the reciprocal lattice vector) are topologically dark\cite{fu_topological_2007}. As is already mentioned, our cleaving surface is therefore a dark surface, and no Dirac cone found in our ARPES data is consistent with the WTI nature. The calculation result of the other type (010) surface is presented in Fig. 4c, and two Dirac points are clearly noticed; see the points indicated by `DP'. The dashed lines with arrows demonstrate that both Dirac points are indeed the points made by the crossings of the surface bands connecting valence and conduction bands, and the band inversion changes the band parity. For another allowed (i.e., non-dark) surface of (100), we found Dirac points as well; see Supplementary Fig. 6. While some Dirac points are embedded in the bulk bands, the Dirac nature of their crossings can always be checked straightforwardly. Hereby, we establish that TaSe$_3$ has a WTI nature. 

It should be noted that the experimental verification of WTI nature can be challenging due to the two distinctive types of topological surfaces and the limited capability of choosing cleaving surface\cite{liu_2016,autes_2016,noguchi_2019}. Further, additional subtleties are also involved in the computation side arising from the notorious issue of  exchange-correlation functionals\cite{liu_2016,autes_2016,noguchi_2019}. Our effort of resolving this issue for the current case of TaSe$_3$ can be found in Supplementary Information. To the best of our knowledge, the only example of the experimentally reported 1D WTI is $\beta$-Bi$_4$I$_4$ and its WTI nature has still been debated until very recently. Our case of TaSe$_3$ is the second, if not the first, example of 1D WTI, and it is the first example of 1D topological superconductor.

In summary, we provide the convincing evidence of WTI nature in TaSe$_3$ by means of  ARPES experiments and first-principles calculations. Topological features of band inversion and surface in-gap state are clearly identified. The calculated $Z_2$ indices as well as the absence and the presence of Dirac cones respectively on the dark and the non-dark surfaces confirm our conclusion of WTI nature. With this, this material is suggested to be the first experimentally confirmed example of q1D topological superconductor in an intrinsic as-is form. Unveiling the relation between the superconductivity and topology can also be an exciting future direction.

\noindent \textbf{Methods}

\noindent \textbf{Sample growth and characterization.}
Single crystals of TaSe$_3$ were grown via the chemical vapor transport (CVT) method\cite{LEVY198361}. Ta (99.99\%) and Se (99.999\%) powder of molar ratio 1\ :\ 3.3 were loaded into one end-side (source zone) of the quartz tube. The additional amount of Se is intended to prevent Se deficiency \cite{1982JSSCh..41..323K}, and to work as a transport agent. Then the quartz tube was evacuated, sealed and loaded into the two-zone furnace with the temperature of the zones set by 720$^\circ$C and 680$^\circ$C, and maintained for 14 days. Whisker-like single crystals of typical dimensions of 10$\times$1.5$\times$0.05 mm$^3$ were achieved. The structural and electrical properties of TaSe$_3$ single crystals were characterized using energy dispersive spectroscopy (EDS), X-ray diffraction (XRD), and electrical resistivity measurement (see Supplementary Fig. 1).\\

\noindent \textbf{ARPES measurements.}
ARPES measurements were performed at beamline 5-4 of Stanford Synchrotron Radiation Lightsource (SSRL), SLAC National Laboratory and beamline 4.0.3 of Advanced Light Source (ALS), Lawrence Berkeley National Laboratory. ARPES spectra were acquired with Scienta R4000 (R8000) electron analyzer at SSRL (ALS).  A clean surface of the sample was obtained by \textit{in-situ} cleaving of single crystals of TaSe$_3$. Samples were cleaved to reveal the natural cleavage plane of (10\={1}) at 10K under ultra-high vacuum pressure better than 5$\times10^{-11}$ torr. Measurements were carried out maintaining sample temperature at 10K at both beamlines. At SSRL, linearly polarized light of photon energy $h\nu$\ =\ 20 eV was used for the measurements. At ALS, linearly polarized light of photon energy $h\nu$\ =\ 56 eV was used. The total energy resolution and angular resolution was set to be better than 10 meV and 0.3$^\circ$, respectively, for the measurements at both beamlines.\\

\noindent \textbf{DFT band structure calculations.}
Our DFT calculations were performed with `Vienna Ab initio Simulation Package (VASP)' based on the projector augmented-wave pseudopotential\cite{kresse_ab_1993,kresse_efficiency_1996}. We used the experimental lattice parameters \cite{wilson_bands_1979}, and the internal atomic coordinates were optimized with a force criterion of 0.001 eV/\AA. The $7\times15\times7$ k-points and the energy cutoff of 600 eV were adopted. SOC was also taken into account. All of the presented  results were obtained within so-called `PBE' generalized gradient approximation (GGA) for exchange-correlational functional  \cite{perdew_generalized_1996}. 
Our main results were double-checked by using GGA-PBEsol \cite{csonka_assessing_2009} and local density approximation (LDA) as parameterized by Perdew and Zunger \cite{ceperley_ground_1980,LDA_PZ_PhysRevB.23.5048}. We also double-checked the results by using all-electron full-potential code, `Wien2k' \cite{schwarz_solid_2003} for which R$_{\rm{mt}}$K$_{\rm{max}}=7$ and $7\times19\times6$ k-points were adopted in the irreducible Brillouin zone. The augmented planewave sphere radii of Ta and Se were 2.48 and 2.36 a.u., respectively.
We used `Wannier90' code to extract MLWFs for Ta $d$ and Se $p$ orbitals \cite{marzari_maximally_1997,souza_maximally_2001,mostofi_wannier90:_2008}, and `WannierTools' code \cite{wu_wanniertools:_2018} to analyze the topological property and surface state. The parities at TRIM points were calculated directly from DFT wave-functions by using `vasp2trace' \cite{vergniory_complete_2019}. For comparison with ARPES data, the $E\mathrm{_F}$ was shifted by 39 meV in the calculated band structure for both bulk and slab case. As reported in previous studies \cite{xia_observation_2009,chen_experimental_2009,hsieh_observation_2009,wang_topological_2016}, it likely reflects the presence of a certain amount of defects in crystals.\\

\noindent \textbf{Acknowledgements}
Y.K. acknowledge helpful discussion with S.-K. Mo and J. S. Kim. M. Y. J. and M.-C.J. thank to S. Nie and Z. Wang for helpful comments. This research was supported by National R\&D Program (No.2018K1A3A7A09056310), Creative Materials Discovery Program (No.2015M3D1A1070672), Basic Science Resource Program (No.2017R1A4A1015426, No.2018R1D1A1B07050869) through the National Research Foundation of Korea (NRF) funded by the Ministry of Science, ICT and Future Planning, 
Basic Science Research Program (No. 2019R1A6A1A10073887) through the NRF funded by the Ministry of Education 
and by the Internal R\&D Program at KAERI funded by the Ministry of Science and ICT (MSIT) of the Republic of Korea (525350-19).
M. Y. J., M.-C. J. and M. J. H. were supported by Basic Science Research Program (2018R1A2B2005204), and Creative Materials Discovery Program through NRF (2018M3D1A1058754) funded by the MSIT of Korea, and the KAIST Grand Challenge 30 Project (KC30) in 2019 funded by the MSIT of Korea and KAIST.\\

\noindent \textbf{Author contributions}
Y.K. and M.J.H. conceived the work. J.H. synthesized and characterized the TaSe$_3$ single crystals. J.H., S.K., Y.L., C.L., J.C., G.L., Y.A., and Y.K. performed the ARPES measurements with support from M.H., D.L. and J.D.D. J.H., S.K. and Y.K. analyzed the ARPES data. M.Y.J., M.-C.J. and M.J.H. carried out theoretical calculations. All authors discussed the results. J.H., M.Y.J., M.J.H. and Y.K. wrote the manuscript with contribution from all authors.

\vspace{10 pt}

\end{document}